\documentclass[aps,prl,preprint,groupedaddress,showpacs]{revtex4}

\usepackage{graphicx}
\begin{document}

\title[Apparent superluminal advancement of a single photon ...]{Apparent superluminal advancement of a single photon far beyond its coherence length}
\author{Simone Cialdi, Ilario Boscolo, Fabrizio Castelli and Vittoria Petrillo}
\affiliation{Istituto Nazionale di Fisica Nucleare and Dipartimento di Fisica, Universit\`{a} di Milano, via Celoria 16, 20133 Milano, Italy}

%\ead{fabrizio.castelli@mi.infn.it}
%\date{\today}

\begin{abstract}
We present experimental results relative to superluminal
propagation based on a single photon traversing an optical system, called
4f-system, which acts singularly on the photon's spectral component
phases. A single photon is created by a CW laser light down--conversion
process. The introduction of a linear spectral phase function will lead
to the shift of the photon peak far beyond the coherence length of the photon
itself (an apparent superluminal propagation of the photon). Superluminal group
velocity detection is done by interferometric measurement of the temporal
shifted photon with its correlated untouched reference. The observed
superluminal photon propagation complies with causality. The operation of
the optical system allows to enlighten the origin of the apparent
superluminal photon velocity. The experiment foresees a superluminal
effect with single photon wavepackets.
\end{abstract}


\maketitle

\section{Introduction}

Einstein's theory of the special relativity establishes
that the velocity $c$ of the light in vacuum is an invariant under a reference
frame change \cite{ein, ein2, bri}. Superluminal single objects under Lorentz
transformations lead to violation of relativistic causality principle,
and in turn, to the paradox of effect preceding its cause
\cite{garrison}. Nevertheless, many experiments with faster-than-$c$ light
propagation were done and discussed \cite{lon,spi,ste,wan}. In fact, special
relativity theoretical framework stands even without assuming that $c$ is the highest
possible speed \cite{ign,leb,lug,recami}.

Research on superluminality refers mostly to barrier tunneling by radiation pulses
\cite{lon,spi} or by single photons \cite{ste}, and to
active media crossing \cite{li}. Barrier tunneling of light pulses is substantially governed by very low transmission coefficients and by an almost linear spectral
component time delay  $\tau_d = d \phi(\omega)/ d \omega$.
Some authors state that the outgoing light pulse after crossing the barrier is so much weaker than the incoming one (or photon crossing probability so scarce) that any possible information carried by the pulse is destroyed, therefore causality principle is not broken down \cite{ste,wan,sol}.
Others \cite{win} maintain that no propagation can occur inside the barrier, hence it is not the case to speak of advancing velocity. In optical pulse propagation experiments within the so-called fast-light media, that is media with anomalous dispersion
\cite{ste,wan} (precisely, with gain-assisted linear anomalous dispersion)
pulse shape is preserved and phase varies almost linearly with frequency in the region
of interest.  The graph slope $d n / d \omega$ leads to a group velocity $v_g$ which exceeds the speed of light in vacuum and can even become negative \cite{wan,ste}. A superluminal experiment carried out with a microwave pulse crossing a birefringent two-dimensional crystal resulted in a clear superluminal group velocity. This was measured using the interference of pulses which had traveled along the two crystal axes
\cite{sol,bru} set in such a way that pulse polarization of the incident and detected fields relative to the crystal fast axis could be well controlled.

The experimental results on these systems renewed the debate
about superluminal propagation and information velocity. The discussion focuses on the concept that the speed of a light pulse crossing a medium is not precisely defined because a pulse is an ensemble of optical components traveling at different and well defined phase velocities $v_p=c/n(\omega)$, where $n(\omega)$ is the refractive index of the optical material at a given frequency. The pulse peak
usually travels at the group velocity $v_g = c/n_g$ where $n_g= n + \omega \,
dn/d\omega|_{\omega=\omega_0}$ is the group index and $\omega_0$ is the
wavepacket central frequency \cite{agrawal}. The wave nature of the
wavepacket allows superluminal light propagation. Arguments concentrate on the fact that $v_g$ does not coincide with the information velocity $v_i$ and there is a debate about the nature of these velocities \cite{garrison,wan,ste}. Since the analysis of the problem by Sommerfield and Brillouin it is discussed that the ``front'' velocity of a square pulse does not exceed $c$, and Refs. \cite{garrison,ste,stenner} suggest a non--analytic point of the pulse amplitude as transporting information, by observing that this is a generalization of the ``front'' point of the pulse.

In all experiments carried out so far, the temporal forward shift of the pulse or of the single photon wavepacket is much smaller than the total length involved, and this necessarily poses interpretation problems. In this respect, the
definition of the information velocity as the propagation speed of a
particular point in the profile \cite{stenner} leads to information
velocities always less or equal to $c$.

Experiments show that the characteristic of a light pulse for providing
superluminal effects is its nature of being a superposition of monochromatic
waves.  Within this view, the possibility of a superluminal effect with a
single photon lies on the fact that a photon is always a superposition of
monochromatic Fock states $|1,\omega\rangle$ (encompassing a frequency
bandwidth due to Heisenberg principle and the fact that the photon is
generated in a definite space region). We would like to underline that in
Quantum Mechanics any single particle is a superposition of many states
(another example, a moving electron is a superposition of momentum
eigenstates $|k\rangle$), even if they are detected in laboratory as a single event
(\emph{i.e.}, a single \emph{click}). Therefore, Quantum Mechanics allows
superluminal propagation of single particles.

We are going to present and discuss an experiment where a single
photon, created in a non-linear crystal by the down-conversion
of a CW laser light \cite{hong}, is operated through its spectral components in such a
way that a clear temporal shift with respect to the non-acted
upon photon is detected via an interferometric measurement. This experiment
pertains to the class of superluminal experiments. The way we carry out
the experiment (acting upon one of the two generated photons only)
does not allow us to claim strictly that we are operating with
one single photon wavepacket. The description of the interference
between optical components can be also done by viewing the observed
light temporal shift as a result of the interference of very weak light beams with femtosecond coherence length. Anyway, for simplicity and convenience, we will use the view of running photons.

In our experiment the photon velocity can become apparently superluminal as a
result of the interference of photon optical components whose phases are
acted upon by the optical system described below. Because of the
interference process, the observed result is not at odds with
causality. In order to avoid issues about the kind and the physical
meaning of the different definitions of velocity, we set the
experiment in such a way that the shift between the normal and
``superluminal'' photons is notably wider than the width of the
corresponding coherence length.

\begin{figure}
\includegraphics[width=11.cm]{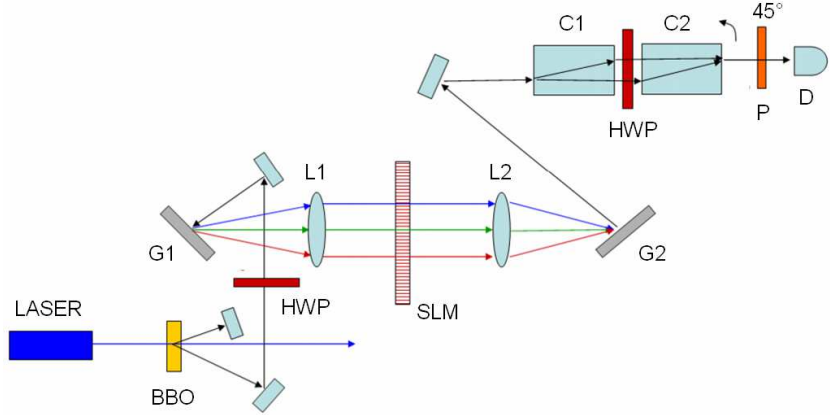}
\caption{\label{f:apparato} Sketch of the experimental apparatus.
Laser: a diode laser of 405 nm wavelength, 40 mW power. BBO: beta-Barium-Borate nonlinear crystal. HWP: half wave plate. G1, G2 : gratings with 1200 lines/mm.
L1, L2 : lenses with f = 100 mm. SLM : a liquid crystal spatial light modulator with 640 pixels. C1, C2 : calcite crystals. P : polarizer oriented at 45$^o$. D : optic coupler + multimode optic fiber + single photon detector.}
\end{figure}

\section{The experiment}
In figure~\ref{f:apparato} we show the experimental device
consisting, in sequence, of a CW pump laser, a non-linear BBO crystal that generates photons via parametric down-conversion, a half-wave-plate (HWP) for
inducing a polarization rotation, a 4f-system with a phase mask (a SLM, \emph{Spatial Light Modulator})
in the middle providing a time delay among optical components of the horizontally polarized
photon beam \cite{wefe,weiner,cialdi}, and finally an interferometer followed by a single photon detector for measuring the time delay.
A pair of photons, usually called \emph{signal} and \emph{idler} photons, is generated by a parametric non-collinear down-conversion \cite{hong} of a CW 40 mW 405 nm almost monochromatic laser radiation within a non-linear 3 mm thick BBO crystal. The state of this photon pair can be written as \cite{joo}
\begin{equation}\label{psi}
    |\Psi_0 \rangle = \int d\omega
    f(\omega)|H, \omega \rangle |H,-\omega \rangle \, ,
\end{equation}
where $\omega$ is the frequency shift with respect to the central frequency $\omega_0$ and $H$ indicates the horizontal polarization.
The function $f(\omega)$, defined in \cite{cia}, gives the probability
amplitude. The signal photon enters the 4f-system, where the required linear delay $\tau_d$ is generated by the mask.
The 4f-system consists of two gratings with 1200 lines/mm and of two 100 mm focal lenses. By means of this device the photon spectral components are spatially dispersed in a linear way by the first grating, and then focused onto the liquid crystal mask array of pixels (our mask is composed by 640 pixels 100 $\mu$m wide) capable of setting the
relative phases almost at will. Finally the optical components are again
synthesized by the second grating. The transmission coefficient of the 4f-system depends on the efficiency of the two gratings, and in our apparatus it is around 50\%. This device is more flexible than the fast-light medium and, more importantly, allows the manipulation of each single optical component.
In this context we observe that our setup allows us to act separately, almost at will, upon the photon optical component phases, contrary to all other experiments in the literature where a dispersion law is imposed by a medium. In our case, the entering light is opened up along a plane by means of the Fourier components' spatial expansion, while in the other
experiments the components were forced to propagate along the same
original line within the acting medium.

Now we analyze the propagation of the signal photon along the
experimental apparatus. A half-wave-plate (HWP) set in front of the the
4f-system rotates the photon polarization of a suitable angle $\theta$ (see below), hence the state of the entering photon is changed into a superposition of a horizontal and a vertical
polarized states:
\begin{equation}\label{state}
    |H, \omega \rangle \rightarrow \cos(\theta) |H, \omega \rangle + \sin(\theta) |V, \omega \rangle \, .
\end{equation}
Then the signal photon propagates through the 4f-system.  The spectral phase function that we introduce is applied only on the horizontal polarization, while the vertical polarization experiences only the
delay due to the transit through the mask pixels, becoming therefore our time
reference. Considering only  the path sections which have different optical thicknesses for H and V polarizations, we obtain the two phase variations:
\begin{equation}\label{boh1}
    \left\{
    \begin{array}{cl}
    \phi_H^m(\omega) \, = & \, (\omega_0 + \omega) \, \tau_H^m \,
  + \, \phi^{SLM}(\omega_0 + \omega) \\
    \vspace{0.1cm}  \\
    \phi_V^m(\omega) \, =& \, (\omega_0 + \omega) \, \tau_V^m
    \end{array}   \right.
\end{equation}
where $\phi^{SLM}(\omega_0 + \omega)$ is the spectral phase function
imposed by mask pixels. In our case we introduce a linear function
$\phi^{SLM}(\omega_0 + \omega) = (\omega_0 + \omega) \, \tau$, with $\tau$ a constant parameter. The times $\tau_H^m$ and $\tau_V^m$ are the time delays due to the pixel slab crossing. We found experimentally that $\Delta\tau^m = \tau_H^m - \tau_V^m = 10$ fs. Incidentally, in the setting of the diagnostic system we also had to
take into consideration the fact that the two transmission
coefficients $t_H$ and $t_V$ of the 4f-system are different (this is
due to the different transmission efficiencies of the gratings for the
two polarizations).

The interferometer placed after the 4f-system to detect the signal photon
at the optical system output is made by two calcite crystals, an HWP
and a polarizer set at 45$^o$. This device, described in \cite{gog}, interchanges the two polarizations and causes a time delay which can be changed simply by rotating the second crystal. We also have to take into account a
certain dispersion introduced by the crystals because they are relatively
long. However this dispersion, described by the parameter $\beta$, can be assumed to be nearly equal for the two paths with a very good approximation. The photon state propagation within crystals is then described by the following spectral phase variation:
\begin{equation}\label{boh} \left\{
    \begin{array}{cc} \phi_H^d(\omega) \, = & \, (\omega_0 + \omega) \, \tau_H^d
      \, +
      \, \frac{1}{2} \, \beta \, (\omega_0 + \omega)^2 \\
      \vspace{0.1cm}  \\
      \phi_V^d(\omega) \, =& \, (\omega_0 + \omega) \, \tau_V^d \, + \,
      \frac{1}{2} \, \beta \, (\omega_0 + \omega)^2 \end{array} \right.
\end{equation}

The last step to be analyzed is the propagation through the polarizer placed
at $45^o$. This is a crucial element in the photon detection.
In fact, the polarizer mixes up the H and V polarization states with the
result that the two states interfere and a pattern of fringes within
the coherence length are created.
Summing up the state evolution along the entire path, the signal component
$|H, \omega \rangle$ at the output read
\begin{eqnarray}
 |H, \omega \rangle  \Longrightarrow  & \frac{1}{\sqrt{2}}
 \left[  t_H \, \cos(\theta) \, e^{i \left( \phi_H^m(\omega)
 + \phi_H^d(\omega)\right)}  \right.  &  \nonumber \\
 & \left.  + \, t_V \, \sin(\theta) \, e^{i \left( \phi_V^m(\omega)
  + \phi_V^d(\omega) \right)} \right] | 45^o, \omega \rangle
  \, = \, A(\omega) | 45^o, \omega \rangle  \hspace{0.3cm} &
\end{eqnarray}

The probability of having one count relative to the signal photon, ignoring
the idler one, is given by  the trace of the density matrix $|\Psi_p \rangle
\langle \Psi_p |$ where:
\begin{equation}\label{psi2}
    |\Psi_p \rangle = \int d\omega f(\omega)
A(\omega)|45^o, \omega \rangle |H,-\omega \rangle \, .
\end{equation}
After some mathematics we get
\begin{eqnarray}
   P(\tau, \Delta \tau) \, & = & \, t^2
   \int d \omega \left| f(\omega) \right|^2 \nonumber \\
   & & \hspace{0.6cm} \times \left[ 1 + Re \left\{
   e^{i \left(\Delta\tau^m + \tau + \Delta \tau \right) (\omega_0 + \omega)}
   \right\} \right]
\end{eqnarray}
where $t = t_H = t_V$ is obtained with a proper rotation of the HWP set in
front of the 4f-system, and $\Delta \tau = \tau_H^d - \tau_V^d$ is the time delay
introduced by the interferometer.

\begin{figure}
\includegraphics[width=11.cm]{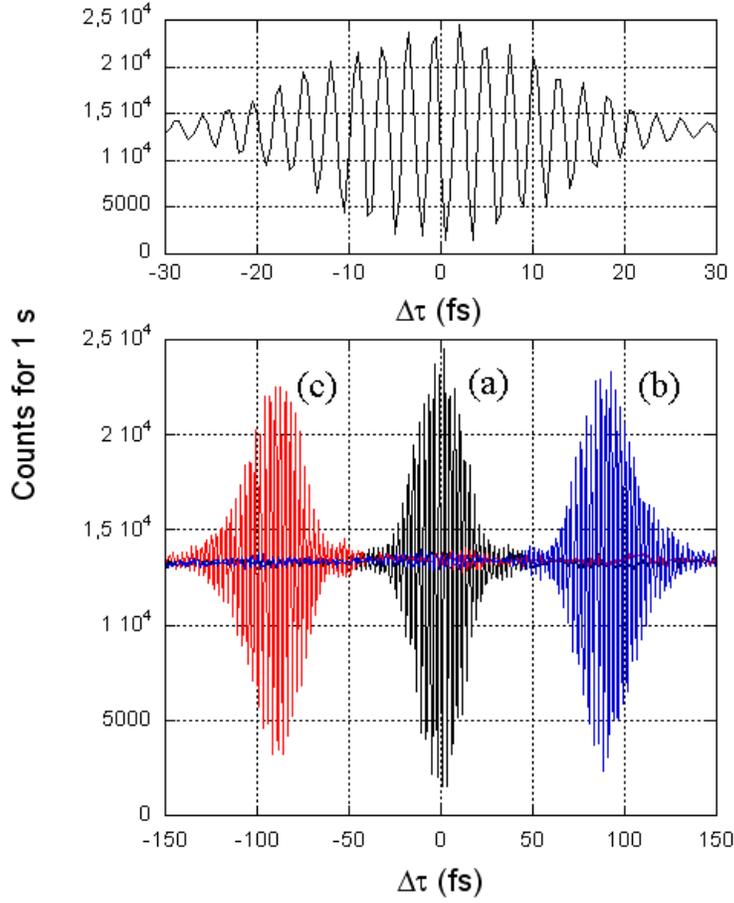}
\caption{\label{f:misure} The interference records of the advanced (b) and retarded (c) photon-wavepackets, referred to the reference $\tau = 0$ (a). The top frame shows an expanded view of the interference fringes.}
\end{figure}

This result accounts for our experimental data shown in
figure~\ref{f:misure}. Curve (a) is the reference case $\tau = 0$ (which sets the origin of the time scale). Curves (b) and (c) present a lead and a lag of $\tau = \mp 100$ fs, respectively. All data show that the interference fringes occur within a coherence length of $30 \, \mbox{fs} < |\tau|$.

\section{Discussion of the experimental results}
Considering the transit time of the photon light from the source up to the end of
the apparatus (that is looking at the optical system as a long black-box), we
observe that the horizontal and vertical polarized parts take
different time intervals. The two parts travel at the same velocity within
the vacuum sections and within the lenses (which are isotropic), hence we may
say that their velocities are different within the mask slab of $\ell_m = 10 \, \mu$m thickness.
Using the delay between V and H states measured at the
end of the apparatus, we may define the group velocity of the
horizontal polarization as
\begin{equation}
v_H^g = \frac{\ell_m}{\ell_m / v_V^g + \Delta \tau^m + \tau}
\end{equation}
where $v_V^g = c/1.488$ is the group velocity of the vertical polarization, derived
from the manufacturer mask characteristics.  The group velocity
defined in this way would result greater than $c$ when $\tau$ is lower than
$-30$ fs and even negative for $\tau < -60$ fs.  We must observe that this
overall view of the light transmission would raise problem with
respect to the causality principle \cite{pea}, because of a photon propagation
mathematically superluminal, and does not consider the real physics of the
phenomenon.

The overall result is readily explained by following the spatially sectioned
sub-light-packets crossing the mask pixels. Each one of these sub-packets has
the limited spectrum selected by the pixel dimension.
That portion of the spectrum corresponds to a coherence length of
3 ps. These sub-packets have subluminal velocity in every part of the device, including the mask. According to this view, the recombination of the sub-packets on the second grating leads to either the forward or backward shift of the photon with respect to the non-acted photon state, depending on the setting of the component phases. This superluminal effect was
already observed in Refs.~\cite{sol, bru}.

The question of the information velocity in our experiment does not fit either the discussion presented so far or the debate in progress about the matter,
that is, within the models of pulse reshaping and consideration of peculiar points of the pulse (such as front or non-analytic ones). In our experiment
photon reconstruction may occur within the entire 3 ps coherence
length of a sub-wavepacket, which means a shift backwards or forwards of the
reconstructed envelope which is very far from the tail of the reference one.
An observation is in order: the delay introduced by the lenses, which is about 30 ps,
is larger than the 3 ps of maximum advance allowed, and this does not allow
direct measurement of ``superluminality'' downstream the 4f-system. However, in
principle, one could substitute the refractive optics with parabolic mirrors
\cite{rei}, this way eliminating the causes of the delay.

The propagation of the spectral sub-light-packets crossing the pixels is
certainly in agreement with causality. In fact, considering Kramers-Kronig
relations for the sub-packets, we can represent the evolution by means of
a Green function that satisfies the requirements of causality. For different
pixels they are independent of one another, so the phase of each pixel can be
programmed at will.
There is no contradiction, then, in saying that the propagation is causal, although the photon moves far ahead its coherent length. From this analysis one can infer that the relevant time is not the coherence time of the photon, but the coherence time of the sub-packet that reaches the single pixel.

\section{Conclusions}
We have performed an experiment on a superluminal shift wider than the coherence length, hence more noticeable than those observed in all other experiments carried out so far. The overall result of figure~\ref{f:misure} indicates clearly that we have induced a large superluminal group velocity on the radiation traveling inside the apparatus.
Our experiment can also be described by considering a single photon propagating within the apparatus. The superluminal group velocity is such that the preserved photon envelope shows up at distances from the vacuum site which are much larger than the photon coherence length, a result that is not possible with the other experimental layouts. This result was obtained with an optical system capable of managing the single component phases of the radiation independently, at variance with all other previous experiments. We have shown that our observations are consistent with the principle of causality even if the nominal group velocity is highly superluminal.

By considering this experiment extendable to single photons, we observe that the results would have a physical content different in essence with respect the complementary result obtained with sub-picosecond laser pulses \cite{lon,spi}. In fact, while the detection of the propagation speed of particular points of a light pulse profile (as for instance the front edge) can be in principle experimentally measured, it cannot be
considered in the case of a single photon because a point within its
wavepacket profile is meaningless, and represents only the probability
amplitude of obtaining a clic (\emph{i.e.} of detecting the photon).
The superluminality experiment with single photons could be carried out neatly thanks to the exploitation of the particular technique of the spatial light modulator which allowed to manage the spectral components while substantially maintaining their amplitudes.

\begin{acknowledgments}
We wish to acknowledge fruitful discussions with C. Maroli and S. Olivares, and the extended help in the experimental work by A. Schiavi. We would like to remark the support by M. Giammarchi. One of the authors thanks V.S. Olkhovsky for stimulating interest in the subject. The research was partly supported by European contract RII3-CTPHI506395CARE.
\end{acknowledgments}

%\section*{References}

\end{document}